\documentclass[12pt,preprint]{aastex}
\newcommand{\mjb}{mJy~beam$^{-1}$}

\newcommand\kms{km~s$^{-1}$}

\newcommand\mcn{\multicolumn{2}{c}{}}

\slugcomment{23 Sept 2006 version}

\shorttitle{Formaldehyde Masers in Sgr B2}
\shortauthors{Hoffman et al.}

\begin{document}

\title{The Formaldehyde Masers in Sgr B2: Very Long Baseline Array and Very Large Array Observations}

\author{Ian M.\ Hoffman}
\affil{St.\ Paul's School, 325 Pleasant Street, Concord, NH 03301}
\email{ihoffman@sps.edu}

\author{W.\ M.\ Goss}
\affil{National Radio Astronomy Observatory, Socorro, NM 87801}

\and

\author{Patrick Palmer}
\affil{Department of Astronomy and Astrophysics, University of Chicago, 5640 S. Ellis Ave., Chicago, IL 60637}

\begin{abstract}
Observations of two of the formaldehyde (H$_2$CO) masers (A and D) in
Sgr~B2 using the VLBA+Y27 (resolution $\approx$ 0\farcs01) and the VLA
(resolution $\approx 9\arcsec$) are presented.  The VLBA observations show
compact sources ($\la 10$ milliarcseconds, $\la 80$~AU) with brightness
temperatures $>10^8$~K.  The maser sources are partially resolved in the
VLBA observations.  The flux densities in the VLBA observations are about
1/2 those of the VLA; and, the linewidths are about 2/3 of the VLA values.
The applicability of a core-halo model for the emission distribution is
demonstrated.  Comparison with earlier H$_2$CO absorption observations and
with ammonia (NH$_3$) observations suggests that H$_2$CO masers form in
shocked gas.  Comparison of the integrated flux densities in current VLA
observations with those in previous observations indicates that (1) most of
the masers have varied in the past 20 years, and (2) intensity variations
are typically less than a factor of two compared to the 20-year mean.  No
significant linear or circular polarization is detected with either
instrument.
\end{abstract}

%\keywords{globular clusters: general --- globular clusters: individual(\objectname{NGC 6397},
\keywords{ISM: individual(\objectname{Sgr~B2}) --- ISM: molecules --- masers --- radio lines: ISM}

\section{Introduction}

The nature of Galactic formaldehyde (H$_2$CO) masers is a growing mystery.
While hundreds of Galactic OH, H$_2$O, and CH$_3$OH masers are known, only
five Galactic star-forming regions have associated H$_2$CO maser emission.
To date, this emission is seen only in the $1_{10}\rightarrow 1_{11}$
transition at 6 cm wavelength.  Shortly after the discovery of the first
H$_2$CO maser in NGC~7538 (Downes \& Wilson 1974; Forster et al.\ 1980), a
radiative pumping model was proposed (Boland and de Jong 1981).  The
H$_2$CO masers discovered subsequently did not meet the conditions required
for this mechanism (Gardner et al.\ 1986; Mehringer, Goss, \& Palmer 1994,
hereafter MGP94; Hoffman et al.\ 2003; hereafter H03).  Thus, twenty-five
years after the discovery of the first H$_2$CO maser, these sources remain
rare and the excitation mechanism remains unknown.

\begin{figure}
\epsscale{0.80}
\rotate
\plotone{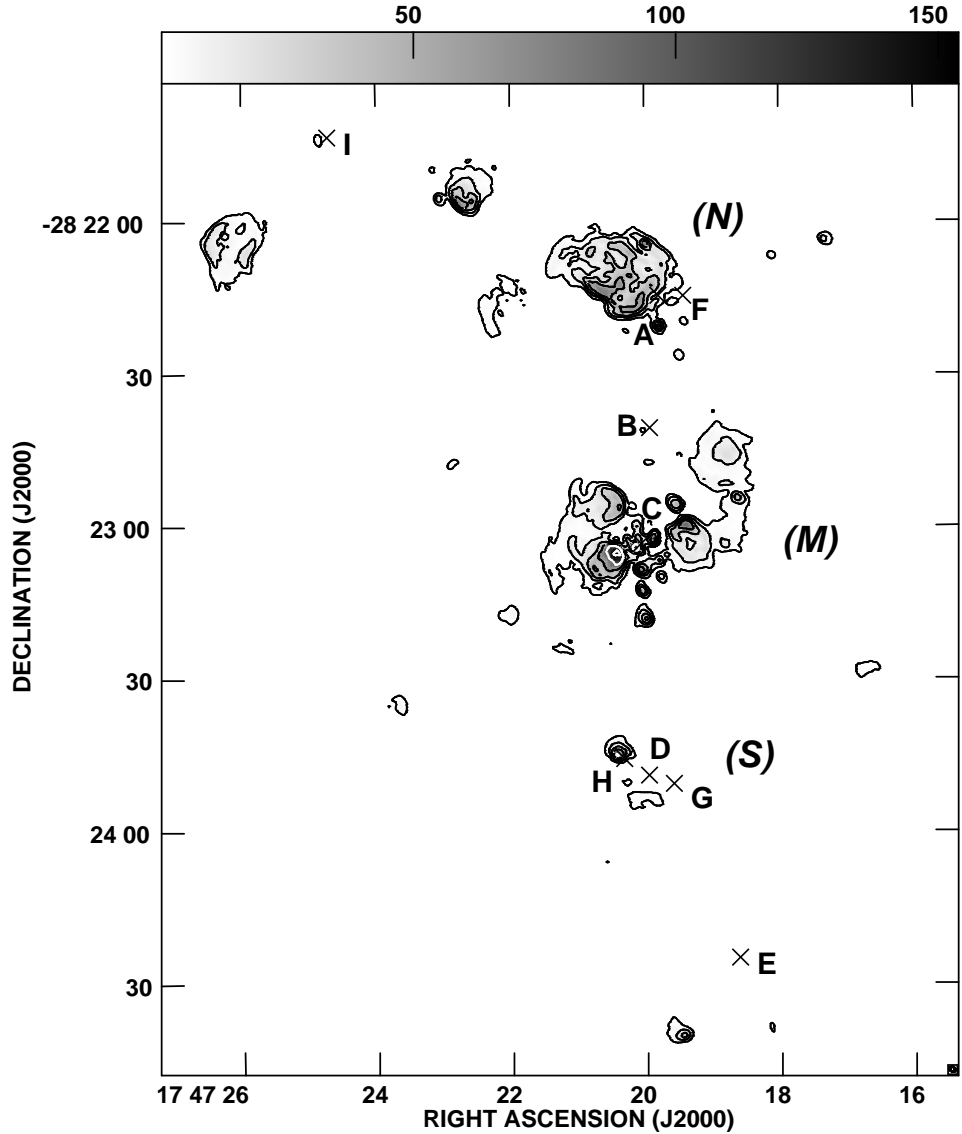}
\caption{Image of the continuum radiation at 4.8~GHz from the Sgr~B2 region from the 1993 VLA-BnA observations (MGP94).
The contours are -6, 6, 24, 64, 128, 256, and 384 times the image RMS noise level of 0.4~\mjb\ and the greyscale at the top is in units \mjb.
The beam is $1\farcs12 \times 0\farcs88$ at a position angle of 52\arcdeg.
The image has not been corrected for attenuation by the primary beam of the VLA antennas (FWHM approximately 10\arcmin).
The labelled crosses indicate the locations of the H$_2$CO maser regions (after MGP94) and the italicized letters in parentheses denote the common designations of the major star-forming complexes.
Masers A and D were detected using the VLBA+Y27.
\label{fig1}}
\end{figure}

Sgr~B2, the northernmost component of the extended Sgr~B radio source, is
located within a few hundred pc of the Galactic center (Reid et al.\ 1988).
(The distance to the Galactic center is assumed to be 8.5~kpc in this
paper.)  Sgr~B2 is comprised of three main star-forming complexes
designated north (N), middle or main (M), and south (S), and many smaller
H{\sc ii} regions.  The H$_2$CO masers occur throughout Sgr~B2, shown in
Figure~\ref{fig1}.  The heating mechanisms and complex chemistry of the
region are subjects of ongoing study (e.g., Gaume \& Claussen 1990;
Goicoechea et al.\ 2004).

Sgr~B2 contains nine individual H$_2$CO maser regions, several of which
have multiple velocity components.  All of the masers are unresolved at
1\arcsec\ angular resolution, except for maser C which MGP94 suggest
consists of several masers blended within the beam.  These regions are near
H{\sc ii} regions distributed over the $\sim$3.6 arcmin$^2$ complex
(MGP94).  Whiteoak and Gardner (1983) and MGP94 designated the maser
regions with letters (Fig.\ \ref{fig1}).  The H$_2$CO masers are observed
over the velocity range $+40$~\kms$\lesssim v_{\rm LSR} \lesssim +80$~\kms,
while other species, such as H$_2$O masers, are observed over a larger
range $-30$~\kms$\lesssim v_{\rm LSR} \lesssim +120$~\kms\ (Kobayashi et
al.\ 1989; McGrath, De Pree, and Goss 2004).  Of the nine maser regions
observed in Sgr~B2 by MGP94, the G maser was shown to be time variable, at
least quadrupling in intensity over 10~yr.  (Similarly, H03 found one
NGC~7538 feature to triple in intensity over $\approx 10$~yr.)  As
initially noted by Whiteoak and Gardner (1983), all of the Sgr~B2 H$_2$CO
masers lie close to OH, H$_2$O, CH$_3$OH, and NH$_3$ masers.  For most of
the masers in MGP94, the separation to an OH maser was less than 0.05 pc.

Recent successes in search techniques for new masers (Araya et al.\ 2004,
2005, 2006a) and in high-resolution observational techniques (H03) promise
to provide empirical constraints for the development of a realistic model
for the Galactic H$_2$CO maser emission.  The necessary steps in compiling
an empirical picture of the H$_2$CO emission in Sgr~B2 are (1) detailed
imaging of the masers in order to quantify the intrinsic properties of the
emission (e.g., brightness temperature), (2) assessment of intensity
variability in the masers, and (3) precise astrometry for elucidating
spatial relationships between the H$_2$CO masers and more common masers
(OH, H$_2$O, CH$_3$OH).  In this paper, we present new observations of the
H$_2$CO masers in Sgr~B2 using the the Very Long Baseline Array (VLBA) and
Very Large Array (VLA) of the NRAO\footnote{The VLA and VLBA are components
of the National Radio Astronomy Observatory (NRAO), a facility of the
National Science Foundation operated under cooperative agreement by
Associated Universities, Inc.}.

\section{Observations and Results}

\subsection{VLBA+Y27}

We observed the A and D H$_2$CO masers in Sgr~B2 using the ten antennas of
the VLBA and the 27 antennas of the VLA as an 11-station VLBI array.
Parameters of the observations are summarized in Table~\ref{tbl-tab1}.  The
total observing time was approximately 8.0 hours alternating between the A
and D pointing positions which are separated by approximately 1\arcmin.  At
each pointing position there is a useful correlated field of view of
approximately 1\arcsec\ (e.g., Bridle \& Schwab 1999).  To observe each of
the nine H$_2$CO masers in Sgr~B2 optimally would have required
observations at nine pointing centers.  Because of signal-to-noise
considerations, we observed only the two masers measured to be most intense
by MGP94.  The baseline lengths of the VLBA+Y27 array range from 52~km to
8611~km; the array is not sensitive to angular scales larger than
0\farcs25.  The antennas have right- (R) and left- (L) circularly polarized
feeds from which RR, LL, RL,
\& LR cross-correlations were formed.  The visibilities were integrated for
8.4~s.  The amplitude scale is set using online system temperature
monitoring and {\it a priori } antenna gain measurements.  The station
delays were determined from observations of J1733-130.

The maser observations were phase referenced to J1745-283 (called W56 in
Bower et al.\ 2001).  Because the properties of this source limit our
observations, we discuss it in some detail.  The absolute position
uncertainty of this source is 12~milliarcseconds (hereafter: mas) (Reid et
al.\ 1999).  Bower et al.\ find that J1745-283 is probably the core of an
extragalactic jet source and that its apparent size at 5~GHz (30~mas) is
determined by scatter-broadening.  We also observed J1745-283 on 22
November 2002 using only the ten VLBA stations (a snapshot observation with
a bandwidth of 4 MHz in both polarizations).  The 2002 observation yields a
deconvolved angular size of $21 \times 18$~mas and an integrated flux
density of 28~mJy; the 2003 observation yields a size of $28 \times 16$~mas
and a flux density of 31~mJy.  (The flux density determined by Y27 [the VLA
alone] during the 2003 observations was 161~mJy.)  Bower et al.\ report an
angular size $42 \times 25$~mas and peak intensity 89~\mjb\ from VLBA+Y1
observations.  Therefore, the flux density must have decreased
significantly between 1999 and 2002.  Such variability would not be unusual
for an inverted spectrum source like J1745-283 (e.g., Urry \& Padovani
1995).

Our imaging of J1745-283 with the VLBA makes use of only the inner-most
stations (VLA, Pie Town, Los Alamos, Fort Davis, Kitt Peak, and Owens
Valley) because the source is resolved on longer baselines.  Some
additional resolution for the maser observations was gained by
self-calibration; but, because only the shorter baselines in the VLBA data
were absolutely calibrated in phase, the position registration accuracy of
the resulting VLBA images is approximately 15~mas.  In summary, we could not
make use of the full potential resolution of the VLBA when using this phase
referencing calibrator, and no other suitable nearby source is known. 

\begin{figure}
\epsscale{0.80}
\rotate
\plotone{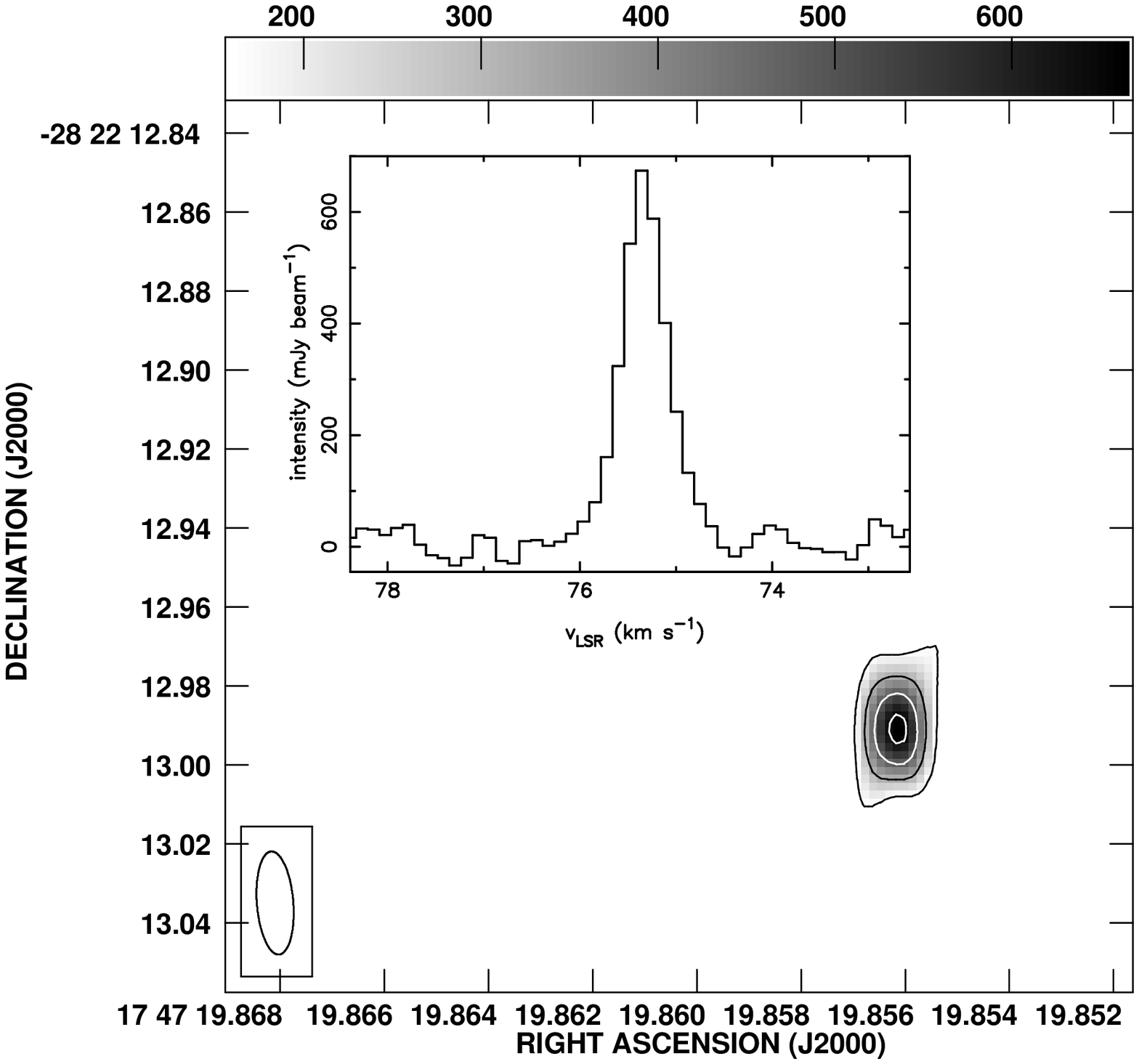}
\caption{VLBA+Y27 image of the H$_2$CO A maser at $v_{\rm LSR} = 75.4$~\kms.
The contour levels are $-$4, 4, 8, 12 and 16 times the image RMS noise level of 40~\mjb\ (no negative contours appear).
The beam (lower left corner) is $26 \times 9$~mas at a position angle of 4\arcdeg.
The greyscale at the top is in units of \mjb.
{\it (inset)} Spectrum at the A maser image peak.
The observed properties are summarized in Table~\ref{tbl-vlba}.
The brightness temperature of the emission is $5.9 \times 10^8$~K.
\label{vlba-a}}
\end{figure}

We detected masers toward the A and D regions.  The F maser is also within
the correlated field of view near the A maser, but was not detected with
the current sensitivity.  The image and spectrum of the A maser from the
VLBI data is shown in Figure \ref{vlba-a}.  Because the radio continuum of
Sgr~B2 is fully resolved by the VLBA, no continuum subtraction was
necessary.  The positions, center velocities ($v_{\rm LSR}$), linewidths
($\Delta{v}_{\rm FWHM}$), peak flux densities ($S_0$), and deconvolved
major and minor axes and position angles are summarized in
Table~\ref{tbl-vlba}.  In parentheses following each entry are the
1-$\sigma$ errors.

The deconvolved sizes of the VLBI images of the A and D masers are $\approx
10$~mas.  At the distance of Sgr~B2, this size corresponds to a linear
diameter of approximately 80~AU, comparable to the sizes (30 to 130~AU)
observed for the H$_2$CO masers in NGC~7538 and G29.96-0.02 using the VLBA
(H03).  The brightness temperatures of the Sgr~B2 A and D masers in the
current VLBI data are $5.9 \times 10^8$~K and $1.2 \times 10^8$~K,
respectively, similar to the $10^{7-8}$~K observed for other H$_2$CO masers
(H03).  No significant linear or circular polarization is detected in
either maser ($\leq$ 20\% for the strongest maser [A]).\footnote{All data
reduction and analysis was performed with the the The Groningen Image
Processing System (GIPSY) software package ({\tt http://www.ast
ro.rug.nl/${\mathtt \sim}$gipsy/}) and the Astronomical Image Processing
System (AIPS) software package ({\tt http://www.nrao.edu/aips/}).}

\subsection{VLA\label{VLA-obs}}

Because of the known variability of H$_2$CO masers (Forster et al.\ 1985; H03), the Sgr~B2 masers were observed in 2005 with the VLA in order to access any possible time-variability.
Parameters of the observations are summarized in Table~\ref{tbl-tab1}.
The array configuration available was DnC for an observation period of approximately 7 hours.
Two pairs of RR and LL bands were recorded centered at the expected velocities of the A and the D masers.

Of the nine maser regions described by MGP94, eight of the sources were detected.
Six velocity components in the regions observed by MGP94 lie outside the velocity range of the current observations.
No new H$_2$CO maser regions were discovered.

\begin{figure}
\epsscale{0.80}
\rotate
\plotone{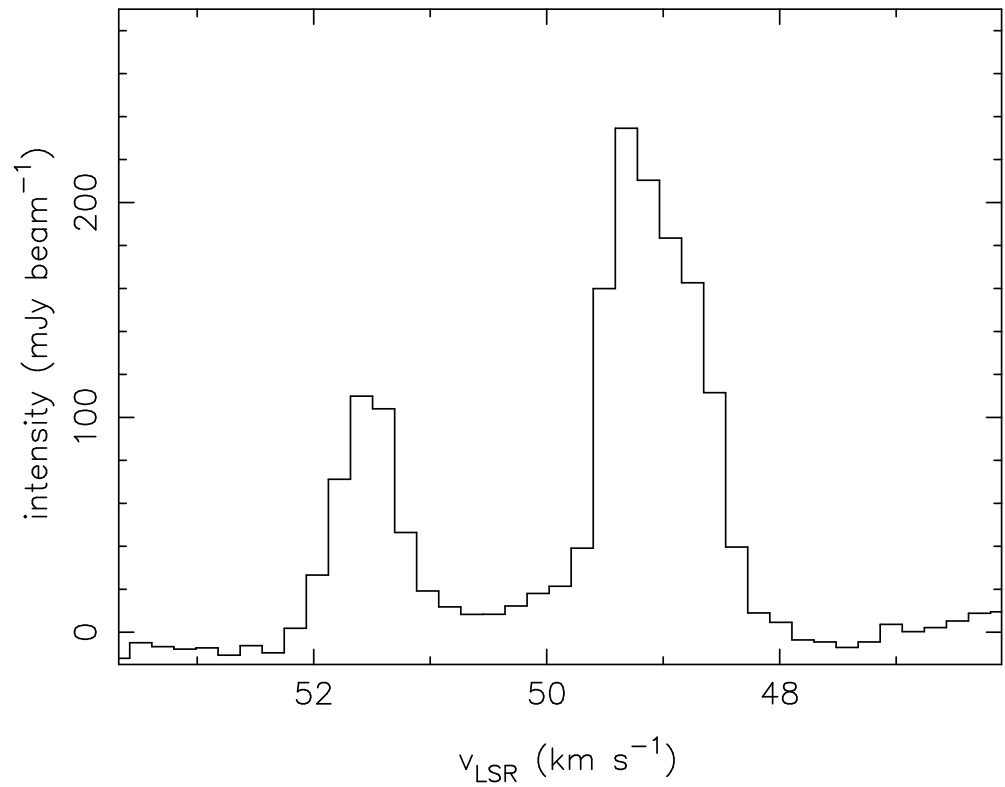}
\caption{VLA spectrum of the E maser from the 2005 observations.
The beam is $10\farcs7 \times 9\farcs0$ at position angle 88\arcdeg\ and the velocity resolution is 0.19~\kms.
\label{vla-e}}
\end{figure}

The 2005 data have inferior angular resolution (10\arcsec\ versus 1\arcsec) but improved spectral resolution (0.19~\kms\ versus 1.5~\kms) compared with the 1993 VLA observations of MGP94.
Comparison of the current data to the MGP94 data is uncertain due to
(1) the severe blending of the strong H$_2$CO absorption with the nearby masers and
(2) the insufficient velocity resolution of the MGP94 data, which did not spectrally resolve the lines.
Variability of the flux density of the masers is apparent and is discussed in \S\ref{disc-vary}.
The 2005 results are summarized in Table~\ref{tbl-vla} and a spectrum from these data (region E) is shown in Figure~\ref{vla-e}.
The C maser region is not discussed in this paper due to confusion of the maser spectra with nearby absorption and continuum emission.
With the current 10\arcsec\ angular resolution of the VLA data, as with the 1\arcsec\ resolution of MGP94, none of the masers are spatially resolved.

In \S\ref{disc-vary}, we also compare the current data with the 1983 VLA observations of Gardner et al.\ (1986).
The velocity resolution of the 1983 data is 0.76 \kms\ and the angular resolution is 1\farcs25.
Although we resolve many of the maser line profiles in velocity for the first time with the 2005 VLA data, most of the linewidths presented in Table~\ref{tbl-vla} agree with the values measured by Gardner et al.\ (1986).

No circular polarization is detected in any of the H$_2$CO masers.
This corresponds to a 3-$\sigma$ upper limit of approximately 4\% circular
polarization for the strongest maser (A).

\section{Discussion}

\subsection{Angular Scatter Broadening}

Images of radio sources are angularly broadened by scattering by the
ionized component of the interstellar medium.  In the direction of the
Galactic center, this problem becomes severe (e.g., Lazio
\& Cordes 1998).  As discussed in \S2.1, J1745-283, is an extragalactic
point-like source whose image is broadened to $\sim$30 mas at 6 cm
wavelength by scattering (Bower et al.\ 2001).  In this section we discuss
the extent to which this scattering medium affects the current VLBI
observations of the H$_2$CO masers in Sgr~B2.

The observed size for J1745-283, $28 \times 16$~mas, is larger than the
observed deconvolved size for the A maser, $13 \times 5$~mas.  This result
may be expected even if both sources are intrinsically point-like because
the proximity of the maser to the scattering medium results in reduced
angular broadening (e.g., Rickett 1990).  Nevertheless, as discussed below,
we expect significant angular broadening in the images of the masers.
Therefore, the deconvolved angular sizes in Table~\ref{tbl-vlba} are upper
limits to the intrinsic sizes and the brightness temperatures are lower
limits.

Gwinn et al.\ (1988) quantified the scattered sizes of H$_2$O and OH masers
in Sgr~B2, finding a $\lambda^{2.2}$ dependence for the broadening.  From
their determinations of minimum angular sizes of 0.3~mas for 1.35-cm H$_2$O
masers and 100~mas for 18-cm OH masers, we expect a minimum apparent size
of approximately 9~mas for the 6.2-cm H$_2$CO masers.  The deconvolved
angular sizes of the masers in Table~\ref{tbl-vlba} are in agreement with
this expectation.

\subsection{Variability of the H$_2$CO masers in Sgr~B2\label{disc-vary}}

In comparing the current VLA observations with earlier VLA data, both the
differing angular resolution and spectral resolution must be considered.
As discussed in \S\ref{VLA-obs}, both the differences in array
configuration and in correlator setup were significant.  To compensate for
the difference in velocity resolution, in Table~\ref{tbl-fd} we tabulate
the velocity-integrated flux density for observations made in 1983 (Gardner
et al.\ 1986), 1993 (MGP94), and 2003 and 2005 (this paper).  However, it
is impossible to compensate for emission outside of the velocity range
covered in the current observations and for the confusion of features
caused by the lower angular resolution.  (The notes in the final column of
Table~\ref{tbl-vla} summarize limitations of the current data from these
causes.)  The error values in Table~\ref{tbl-fd} are dominated by
systematic uncertainties rather than by thermal noise in most cases.

We find that the A maser has approximately doubled in velocity-integrated
flux density since 1993, while B, G, and H increased between 1983 and 1993,
and subsequently decreased.  The velocity-integrated flux densities from
regions D, E, F, and I have not changed significantly since 1993.

\subsection{Core/Halo Morphology}

For the A and D H$_2$CO masers in Sgr~B2, a comparison of the current
VLBA+Y27 data with past and current VLA data shows two major differences:
(1) only a fraction of the flux density seen with the VLA is observed in
the VLBI data ($\sim$70\% for the A maser; $\sim$40\% for the D maser); and
(2) the velocity widths in the VLBI data are more narrow than in the VLA
data ($\sim$80\% for the A maser; $\sim$50\% for the D maser).  In this
section we discuss a schematic model for the structure of the emission
which addresses both the observed angular distribution and velocity widths.

The flux density detected by the VLA but resolved by VLBI baselines must
lie at angular scales between $\approx 100$~mas (the resolution of the
shortest VLBA+Y27 interferometer spacings) and $\approx 500$~mas (the
masers are unresolved in the MGP94 observations with 1\arcsec\ resolution),
while the flux density observed with the VLBI is emitted by a source
$\sim$10~mas.

These limits suggest a core-halo source morphology.  A similar morphology
was proposed by H03 for the H$_2$CO masers in NGC~7538 and G29.96-0.02.  A
model consisting of two coincident circular Gaussian components with
different angular sizes can reproduce the observed results.  For the A
maser, the flux density observed with the VLA is about 2000~mJy.  To
determine the decrease in flux density with increasing projected baseline
length, we use the highest signal-to-noise baselines only: those between
Y27 and other antennas.  The average flux densities observed on baselines
with Y27 are: 1400~mJy at 0.71~M$\lambda$ (to Pie Town, average projected
spacing $\sim$44~km)\footnote{M$\lambda$ denotes million wavelengths}; 1300~mJy at 3~M$\lambda$; and 1200~mJy at
5~M$\lambda$.  These data are fit by a two component Gaussian model with a
600~mJy in a 300~mas halo and 1400~mJy in a 10~mas core.  The brightness
temperature of the halo is $T_B \approx 4 \times 10^5$~K; for the 10 mas
core component, $T_B \approx 8 \times 10^8$~K (Table~\ref{tbl-vlba},
Fig.~\ref{vlba-a}).  A more precise model is justified only after higher
signal to noise measurements.

It is reasonable to assume that the background radiation is relatively
uniform on 300~mas angular scales because the continuum emission is well
resolved on the VLBA baselines.  Therefore, the difference in brightness
temperatures between the core and halo components may be attributed to
differences in maser gain.  We estimate that the halo has a gain of
approximately 10 (in the exponential amplification regime) and the core has
a gain of approximately 2000 (in the saturated regime).  Because the
observed angular sizes are upper limits for the intrinsic sizes, the
derived brightness temperatures are lower limits for the intrinsic
brightness temperatures; therefore, the actual gains may exceed the above
values.  The gain of the maser medium is dependent upon four factors: the
pathlength, the density of H$_2$CO, the level inversion, and the velocity
coherence (see \S4.2 of H03).  Therefore, one or more of these factors must
be significantly different between the core and halo components.  We
suggest that the narrower line widths arising from the core component
indicate that the velocity coherence is enhanced in the core region
compared to the halo, yielding a higher gain.  Similar arguments may be
applied to the D maser, but the resulting ranges of size and $T_B$ are not
well constrained because of the higher noise level on the VLA-Pie Town
baseline.

Additional maser species in other star-forming regions -- excited OH
(Palmer, Goss, \& Devine 2003), OH in supernova remnants (Hoffman et al.\
2005), as well as other H$_2$CO masers (H03) -- exhibit a similar narrowing
of line widths between VLA and VLBA angular scales.
%spell check to here
\subsection{Associations with Other Molecular Lines}

H$_2$CO absorption toward Sgr~B2 has been studied extensively with
10\arcsec \ $\times$ 20\arcsec \ resolution both by Mart\'{\i}n-Pintado et
al. (1990) and Mehringer et al. (1995).  Mart\'{\i}n-Pintado et al. (1990)
noted that the absorption is dominated by three velocity components
($v_{\rm LSR}$= 55, 64, and 80 \kms), each with $\tau>1$.  Linewidths of
these absorption features ranged from 9 -- 26 \kms, i.e. much greater than
that of the maser features ($< 1$ \ \kms; see Table~\ref{tbl-vla}).
Mehringer et al. (1995) provide optical depth profiles toward selected
positions.  All but maser E lies within 30\arcsec \ of one of these positions
displayed.  (Maser E lies at a position with no radio continuum.)  It is
striking that the maser velocities, except for maser I (which is more than
a beamwidth away from the position of a displayed profile), do not occur
in velocity ranges with large H$_2$CO optical depths.  Masers A and F occur in
a velocity range with $\tau$=1, but this velocity range is a rather sharp
minimum in the optical depth profile at this position.  Therefore, we
conclude that the gas containing the H$_2$CO masers is distinct from the
bulk of the H$_2$CO containing gas observed in absorption.

Insight into why the maser containing gas may be distinct is provided by
(3,3) and (4,4) NH$_3$ observations made with the VLA by Mart\'{\i}n-Pintado
et al. (1999).  With their 3\arcsec \ resolution, 80-90\% of the NH$_3$
emission is resolved out, and only small angular scale features remain.
Among these features are a number of rings and arcs, most naturally
interpreted as complete or partial shells with linear sizes $\sim$2~pc.  The
shells are hot (T$_k \sim$ 50 -- 70 K), and the H$_2$ densities derived for
them are typically a factor of 10 greater than those derived from the
H$_2$CO absorption studies of Mart\'{\i}n-Pintado et al. (1990).  H$_2$CO
masers C, D, E, G, and H fall within the field imaged by Mart\'{\i}n-Pintado
et al. (1999).  As these authors note, all of the H$_2$CO masers occur at
positions on hot NH$_3$ shells.  The higher temperature (T$_k >$ 100 K)
region of apparent interaction between shells A and B contains the closely
spaced D, G, and H masers.  Mart\'{\i}n-Pintado et al. (1999) propose that
the pumping mechanism for the H$_2$CO masers as well as that for the NH$_3$
and class II CH$_3$OH masers must depend on the physical conditions in the
hot shells.

In the search for additional constraints on the H$_2$CO emission
environment, we examine the possible association of H$_2$CO masers with
other molecular emission for which physical conditions are better
understood.  In Table~\ref{tbl-others} we present a summary of possible
associations with H$_2$O masers observed by McGrath, Goss, and De Pree
(2004), CH$_3$OH masers observed by Houghton and Whiteoak (1995), and OH
masers observed by Argon, Reid, and Menten (2000).  The excitation
conditions for the different species are mutually exclusive, but if the
masers exist in a shocked region as proposed by Mart\'{\i}n-Pintado et
al. (1999) for Sgr B2 (and for NGC 7538 by H03), a rapid change in
densities, temperatures, and velocity fields over a small linear distance
is to be expected.  Of the NH$_3$ masers in Sgr~B2 with interferometically
determined positions, only masers M1 and M6 lie similarly near an H$_2$CO
maser (E), and both differ in velocity by $>$10 \kms\ from the H$_2$CO
maser.
The existence of H$_2$O and CH$_3$OH masers near the H$_2$CO F maser is
significant because previously the F maser had no known maser or continuum
associations (MGP94).

The H$_2$CO maser in NGC~7538 (H03) and many of those reported in Section
3.2 varied in intensity with timescales from years to decades.  H03 noted a
common variability of some of the H$_2$CO and H$_2$O masers in NGC~7538.
This was further confirmed by long-term observations by Lekht et al.\
(2003, 2004b).  Similarly in Sgr~B2, the long-term monitoring of the H$_2$O
masers presented by Lekht et al.\ (2004a, 2004c) may indicate common
variability of H$_2$O masers with the H$_2$CO A maser.  However,
variablity of an H$_2$CO maser in IRAS 18566+0408 was recently seen to
occur on much more rapid timescale (Araya et al. 2006b).

\section{Conclusions}

We present VLBA+Y27 images of the Sgr~B2 A and D H$_2$CO masers.  The
measured sizes ($\la 80$~AU) and brightness temperatures ($>10^8$~K) are
comparable to those found in other VLBI studies of H$_2$CO masers.
However, about half of the flux density from these regions is resolved out
with the VLBA data.  A comparison between VLA and VLBA observations shows
that the missing flux density exhibits a broader linewidth than the
emission from the compact VLBI source.  We demonstrate quantitatively the
applicability of a core-halo model for these masers.

We also present new VLA observations of the H$_2$CO masers in Sgr~B2 with
improved velocity resolution.  We have detected variability in several of
the masers.

H$_2$O and CH$_3$OH masers discovered near the H$_2$CO masers may indicate
associations among the species, suggestive of a related origin.  The
association of H$_2$CO masers in Sgr~B2 with hot NH$_3$ shells proposed by
Mart\'{\i}n-Pintado et al. (1999), together with the arguments for shock
excitation of the maser region in NGC~7538 in H03, provide an
encouraging stepping stone toward a solution of the problem of the
excitation of these poorly understood masers.
\acknowledgments
We are indebted to an anonymous referee for a very careful reading of this
paper and for directing our attention to two very important references.

%We thank B.\ G.\ Clark and J.\ M.\ Wrobel for flexible scheduling of the Very Large Array.

{\it Facilities:} \facility{VLA ()}, \facility{VLBA ()}.

\clearpage

\begin{deluxetable}{l c c}
%\tabletypesize{\scriptsize}
%\rotate
\tablecaption{Observational Parameters\label{tbl-tab1}}
\tablewidth{0pt}
\tablehead{
\colhead{Instrument} & \colhead{VLBA+Y27} & \colhead{VLA}
}
\startdata
Observing date(s) & 2003 May 17 \& 22 & 2005 October 11, 13, \& 14 \\
Position (J2000.0)& 17 47 20.0463, -28 23 46.587 & 17 47 19.96, -28 22 59.8  \\
         & 17 47 19.8562,  -28 22 12.900 & \\
Synthesized beam  & (A) $20 \times 15$ mas, P.A.= $1{\arcdeg}$ & $10\farcs7 \times 9\farcs0$, P.A.= $88{\arcdeg}$ \\
                  & (D) $35 \times 21$ mas, P.A.= $1{\arcdeg}$ &                                                  \\
Flux density calibrator & --- & 3C286 \\
Phase calibrator & J1745-283 & J1751-253 \\
Bandpass calibrator & J1733-130 & J1733-130 \\
Rest frequency & 4829.6569~MHz & 4829.6590~MHz \\
%Number of IF's & 1 & 4 \\
Number channels & 128 & 63 \\
Channel spacing & 1.953 kHz & 3.052 kHz \\
Velocity Resolution\tablenotemark{a} & 0.24 \kms & 0.19 \kms \\
Center velocity & 50.6 \kms\ \& 75.6 \kms \ & 51.4 \kms\ \& 74.6 \kms \\
Total velocity range & $\pm$7.5 \kms  & $\pm$5.5 \kms  \\
Typical noise per channel &   35 \mjb & 5 \mjb \\
\enddata
\tablenotetext{a}{after Hanning smoothing}
\end{deluxetable}

\clearpage

\begin{deluxetable}{c r@{ }r@{ }r@{.}l r@{ }r@{ }r@{.}l c c c c c c}
%\tabletypesize{\scriptsize}
\rotate
\tablecaption{VLBA+Y27 H$_2$CO Results from 2003 May\label{tbl-vlba}}
\tablewidth{0pt}
\tablehead{
\colhead{Source} & \multicolumn{4}{r}{$\alpha$ (J2000)} &
\multicolumn{4}{r}{$\delta$ (J2000)} & $v_{\rm LSR}$ & $\Delta{v}_{\rm FWHM}$ & $S_0$ & $\theta_a$ & $\theta_b$ & P.A. \\
 & \multicolumn{4}{r}{} & \multicolumn{4}{r}{} & (\kms) & (\kms) & (\mjb) &
mas & mas & $\deg$
}
\startdata
A & 17&47&19&856(1) & $-$28&22&12&99(2) & 75.33(1) & 0.59(2) & 645(15) & 13(3) & 5(2) & 110(20) \\
D &   &  &20&047(2) &      &23&46&59(2) & 50.07(2) & 0.36(5) & 160(20) & 10(4) & 8(4) & 5(40) \\
\enddata
%\tablecomments{RA, Dec}
\end{deluxetable}

\clearpage

\begin{deluxetable}{c r@{ }r@{ }r@{.}l r@{ }r@{ }r@{.}l c c c c}
\tabletypesize{\scriptsize}
%\rotate
\tablecaption{VLA H$_2$CO Results from 2005 October\label{tbl-vla}}
\tablewidth{0pt}
\tablehead{
\colhead{Source} & \multicolumn{4}{r}{$\alpha$ (J2000)} & \multicolumn{4}{r}{$\delta$ (J2000)} & $v_{\rm LSR}$ & $\Delta{v}_{\rm FWHM}$ & $S_0$ & Notes \\
 & \multicolumn{4}{r}{} & \multicolumn{4}{r}{} & (\kms) & (\kms) & (\mjb) & \\
}
\startdata
A & 17&47&19&94 & $-$28&22&13&0  & 75.31(7)& 0.71(2) &1900 & U \\
B &   &  &20&04 &      &22&40&6  & 51.0(2) & 0.4(2)  &  80 & U \\
D &   &  &20&01 &      &23&47&2  & 50.1(1) & 0.75(5) & 510 &  \\
  &   &  & \mcn &      &  & \mcn & 53.9(1) & 0.88(9) & 320 &  \\
E &   &  &18&64 &      &24&24&5  & 49.1(1) & 0.94(4) & 230 & F \\
  &   &  & \mcn &      &  & \mcn & 51.5(1) & 0.60(6) & 120 & F \\
F &   &  &19&61 &      &22&13&5  & 76.3(2) & 0.3(1)  &  40 & U \\
G &   &  &19&57 &      &23&49&9  & 48.7(3) & 0.8(2)  & 120 & U \\
H &   &  &20&43 &      &23&46&7  & 70.6(3) & 0.3(2)  &  50 & U \\
  &   &  & \mcn &      &  & \mcn & 73.1(1) & 0.9(5)  &  50 & U \\
  &   &  & \mcn &      &  & \mcn & 74.6(1) & 0.8(3)  &  70 & U,N \\
I &   &  &24&72 &      &21&43&0  & 70.9(1) & 0.92(6) &  60 &  \\
\enddata
\tablecomments{\\
F -- full velocity range of features observed earlier not covered \\
N -- possible new velocity feature\\
U -- position, velocity, and/or intensity uncertain due to blending (see \S\ref{VLA-obs})
}
%\tablenotetext{a}{After Table 2 from MGP94.}
\end{deluxetable}

\clearpage

\begin{deluxetable}{c c c c c}
%\tabletypesize{\scriptsize}
%\rotate
\tablecaption{Integrated Flux Density Comparisons 1983 --- 2005  \label{tbl-fd}}
\tablewidth{0pt}
\tablehead{
\colhead{Source} & \multicolumn{4}{c}{Integrated Flux Density (mJy \kms) } \\ \cline{2-5}
 & 1983\tablenotemark{a} & 1993\tablenotemark{b} & 2003\tablenotemark{c} & 2005\tablenotemark{c} \\
 & (VLA) & (VLA) & (VLBA) & (VLA)
}
\startdata
A & 720$\pm$15 & 850$\pm$15 & 640$\pm$50 & 1900$\pm$400 \\
B &  80$\pm$15 & 150$\pm$10 & ---        &   20$\pm$10 \\
D & 700$\pm$25 &1400$\pm$30 &  55$\pm$20\tablenotemark{d} & 1200$\pm$300 \\
E & 230$\pm$80 & 350$\pm$20 & ---        &  330$\pm$30 \\
F &$<60$       &  70$\pm$15 & ---        &   40$^{+40}_{-20}$ \\
G &$<30$       & 320$\pm$15 & ---        &   75$^{+50}_{-25}$ \\
H &140$\pm$60  & 350$\pm$15 & ---        &   35$\pm$20 \\
I &$<60$       & 100$\pm$15 & ---        &   75$\pm$20  \\
\enddata
\tablenotetext{a}{from Whiteoak and Gardner 1983}
\tablenotetext{b}{from MGP94}
\tablenotetext{c}{this paper}
\tablenotetext{d}{Includes only one of the velocity components detected in the VLA observations}
\end{deluxetable}

\clearpage

\begin{deluxetable}{c c c c}
\tabletypesize{\scriptsize}
%\rotate
\tablecaption{Nearby Masers \label{tbl-others}}
\tablewidth{0pt}
\tablehead{
\colhead{Source} & \multicolumn{3}{c}{Displacement from H$_2$CO Maser} \\ \cline{2-4}
                 & $\Delta{\theta}$\tablenotemark{a} &
                 $\Delta{\ell}$\tablenotemark{b} & $\Delta{v_{\rm LSR}}$ \\
                 & (\arcsec) & (AU) & (\kms) }
\startdata
\cutinhead{H$_2$O}
A & 0.6 & 5000 &  0.1 \\
  & 0.1 &  850 & 18.8 \\
F & 0.2 & 1700 &  9.9 \\
H & 0.7 & 5500 &  6.6 \\
  & 0.7 & 5500 & 11.9 \\
\cutinhead{CH$_3$OH}
B & 0.7 & 6000 & \tablenotemark{c} \\
D & 0.8 & 7400 & \tablenotemark{c} \\
E & 0.5 & 4200 & \tablenotemark{c} \\
F & 4.5 &38000 & \tablenotemark{c} \\
I & 0.5 & 4200 & \tablenotemark{c} \\
\cutinhead{OH\tablenotemark{d}}
A & 0.6 & 5000 & 2.5 \\
B & 0.6 & 5000 & 1.2 \\
C & 0.15&  130 & 0.2 \\
  & 0.05&   40 & 0.2 \\
  & 0.06&   50 & 0.2 \\
  & 0.18&  150 & 0.2 \\
D & 0.07&   60 & 0.9 \\
  & 0.10&  850 & 1.1 \\
H & 0.21&  180 & 0.1 \\
  & 0.24&  200 & 2.0 \\
\enddata
\tablenotetext{a}{angular separation}
\tablenotetext{b}{projected linear separation}
\tablenotetext{c}{numerous velocity components}
\tablenotetext{d}{Zeeman shifting makes velocity comparison uncertain}
\tablerefs{McGrath, Goss, \& De Pree (2004); Houghton \& Whiteoak (1995); Argon, Reid, \& Menten (2000)}
\end{deluxetable}

\clearpage

\clearpage

\clearpage

\end{document}